\documentclass{article}
\usepackage{amsmath}
\usepackage{amssymb}
\usepackage{amsthm}
\usepackage{mathtools}
\usepackage{float}
\usepackage{empheq}
\usepackage{graphicx}
\usepackage[colorlinks=true, allcolors=blue]{hyperref}
\theoremstyle{plain}
\newtheorem{theorem}{Theorem}[subsection]

\theoremstyle{remark}

\usepackage{booktabs}
\usepackage[letterpaper,top=2cm,bottom=2cm,left=3cm,right=3cm,marginparwidth=1.75cm]{geometry}

\providecommand{\subjclass}[2][]{%
  \par\noindent\textbf{Mathematics Subject Classification (#1).} #2\par%
}
\title{A Quantile-Based Framework for Bimodal Distribution Construction}

\author{
Dreamlee Sharma\thanks{corresponding author: dreamleesharma@yahoo.in}  \\
University of Delhi \\
\and
Nikita Rawat\\
University of Delhi
}

\begin{document}
\maketitle
\begin{abstract}
This study suggests a novel quantile-based framework to construct bimodal distribution which does not rely on mixture representations or closed-form probability density functions. This proposed approach induces bimodality by transforming the quantile density through a strictly positive modulation function so that it can preserve the support and tail behavior of a chosen baseline distribution with providing explicit control over modal structure. The method is shown to apply broadly across bounded, semi-bounded and unbounded distribution families which include power, reflected power, logistic, Weibull and generalized lambda distributions. Simulation studies and a real-data application were performed which have shown that the proposed construction yields smooth, interpretable and computationally stable bimodal models and it also provides a flexible alternative to classical mixture-based approaches. Also this framework allows gradual transitions between unimodal and bimodal shapes which make it well suited for modeling overlapping regimes which are commonly encountered in real-world data. This proposed methodology therfore offers a robust and interpretable tool for bimodal modeling without requiring latent class assumptions or complex estimation procedures.
\end{abstract}
\subjclass[2020]{Primary 62G05; Secondary 62E10, 62P99, 60E05}
\noindent \textbf{Keywords:}Quantile-based distributions, Bimodality, Generalized lambda distribution, Distributional deformation, Mixture-free modeling

\section{Introduction}
Quantile functions have provide a fundamental alternative to probability density  functions for characterizing random variables. In this approach we do not require the likelihood directly as the quantile-based approach models distributions through the inverse cumulative distribution function and hence it offer direct control over location, scale, skewness, and tail behavior. Early work by Parzen \cite{parzen1979} has emphasized on the importance of quantile representation in statistical modeling after this Gilchrist\cite{gilchrist2000} has established quantile functions as a complete and flexible framework for distributional analysis. Recently Prendergast and Staudte \cite{prendergast2019} has formalized the class of quantile-based distributions which highlights their theoretical appeal and practical relevance  particularly when closed-form densities are unavailable. Bimodality have played a central role in statistical modeling as it naturally arises in heterogeneous populations, biological processes, reliability studies and economic data. The most widely adopted approach for modeling bimodal data  relies on finite mixture distributions where the observed density is expressed as a weighted combination of simpler unimodal components \cite{mclachlan2000}. Mixture models have offer flexibility but offer few challenges which includes identifiably issues, sensitivity to initialization, interpretational ambiguity and computational instability \cite{titterington1985, fruhwirth2006}. These limitations motivate the need of alternative mechanisms for concept of multimodality  representations. Finite mixture model approach is used recently for modeling bimodal data but recent work has shown that bimodality can be generated through smooth distributional
deformations and quantile-based constructions without invoking latent class structures\cite{elal2018,gomezdeniz2019, prendergast2019}.
An alternative approach to bimodality is transformation and reflection techniques. Reflection-based constructions where a random variable is mapped to its complement or mirror image have been employed to reverse skewness or induce asymmetry while preserving bounded support \cite{johnson1995}. Such techniques are particularly common in bounded distributions which includes the power and beta families and have been used in reliability and lifetime modeling contexts \cite{lawless2003}. While reflection alters the orientation of the distribution it does not introduce new modes and hence can not generate controlled multimodality. 

A notable advantage of the quantile-based paradigm is its ability to accommodate a wide range of distributional families through explicit quantile functions. Classical distributions such as logistic, Weibull, exponential, power and generalized lambda distributions all admit closed-form or numerically tractable quantile representations \cite{hosking1997, johnson1994, kotz2000}. These properties make the quantile framework particularly attractive to develope unified modeling strategies that transcend specific parametric forms. The need for bimodal distributions extends beyond descriptive modeling. In practice, bimodality may represent competing failure mechanisms, latent subpopulations, or regime-switching behavior. Traditional density-based techniques often obscure this relationship whereas quantile-based modeling offer a natural pathway for linking distributional shape to  structural features. In this paper, we have proposed a novel quantile-density modulation framework to construct bimodal distributions which rather than combining multiple distributions or  by using  ad hoc transformations to introduce bimodality by modifying the quantile density through a strictly positive modulation function. This approach preserves the underlying support and tail behavior of the baseline distribution with the introduction of precise controlled modal structure. We have also established some theoretical conditions under which bimodality exists and demonstrate how modal locations can be explicitly governed by the parameters of the modulation function. The proposed method can be applied uniformly across bounded, semi-bounded and unbounded distributions which includes power, reflected power, logistic, Weibull and generalized lambda families. We have further validated the proposed method through simulation studies and real-data applications which demonstrates that the bimodality arises structurally from the quantile-density formulation rather than from mixture representations. Our results have clearly shown that the method yields smooth, interpretable and computationally stable bimodal models without requiring closed-form probability densities. Thus, this work has brought advancement in the quantile-based modeling literature by using a unified and theoretically grounded mechanism for controlled bimodality.

\section{Methodology}
Let $X$ be a continuous random variable with cumulative distribution function  $F$.  
The quantile function is defined as
\[
Q(p) = F^{-1}(p), \qquad 0<p<1,
\]
and provides an equivalent characterization of the distribution \cite{parzen1979, gilchrist2000}.  
We assume that $Q(p)$ is absolutely continuous on $(0,1)$ and has derivative
\[
q(p) = Q'(p),
\]
which is referred to as the \emph{quantile probability density function} (QPDF) \cite{gilchrist2000, prendergast2019}.  
The quantile function admits the integral representation
\[
Q(p) = Q(0^+) + \int_0^p q(u)\,du,
\]
where positivity of $q(p)$ ensures monotonicity of $Q(p)$ and hence the validity of the induced probability distribution, whenever a probability density function (PDF) exist then the PDF and QPDF are related through
\[
f(Q(p)) = \frac{1}{q(p)}, \qquad 0<p<1,
\]
a result that follows from differentiation of the identity $F(Q(p))=p$ \cite{parzen1979}.  
This relationship implies that \emph{local maxima of the PDF correspond to local minima of the QPDF}.  
Consequently we can say the number and location of modes of the distribution are governed by the critical points of $q(p)$ a principle consistent with classical results in shape analysis and critical point theory \cite{silverman1981, hartigan1985, morse1934}.

\subsection{Quantile-density modulation}
Let $Q_0(p)$ denote a baseline quantile function corresponding to a unimodal distribution and has the associated QPDF written as
\[
q_0(p) = Q_0'(p) > 0 \qquad 0<p<1.
\]
Following the quantile-based distributional framework of \cite{gilchrist2000, prendergast2019} bimodality is introduced by defining a modified QPDF of the form
\[
q(p) = q_0(p)\,h(p)
\]
where $h(p)$ is a strictly positive modulating function.  
The resulting quantile function is then given by
\[
Q(p) = Q(0^+) + \int_0^p q_0(u)h(u)\,du
\]
This construction preserves absolute continuity and monotonicity of the quantile function while allowing direct control over the modality of the distribution.

\subsection{Assumptions on the modulating function h(p)}
To ensure that the resulting distribution is well-defined and bimodal the following assumptions are imposed on $h(p)$:
\begin{itemize}
\item[(H1)] \textbf{Positivity:} $h(p) > 0$ for all $p \in (0,1)$.
\item[(H2)] \textbf{Smoothness:} $h(p) \in C^1(0,1)$.
\item[(H3)] \textbf{Bimodality condition:} $h(p)$ possesses exactly two local minima in $(0,1)$.
\end{itemize}

Under these conditions one can say that the modified QPDF $q(p)$ will have exactly two local minima which implies that the induced PDF will have exactly two modes.This derivative-based control of modality when aligned with established results in shape-constrained inference and density modality theory \cite{silverman1981, robertson1988}.

\subsection{General form of h(p)}
A general and flexible class of modulating functions satisfying assumptions (H1)--(H3) is given by
\[
h(p)
=
c
+
\sum_{j=1}^{2}
a_j
\phi\!\left(
\frac{p-\xi_j}{\delta_j}
\right)
\qquad 0<p<1
\]
where $c>0$ ensures strict positivity, $a_j>0$ control the depth of the modulation, $0<\xi_1<\xi_2<1$ determine the locations of the modes and $\delta_j>0$ regulate their sharpness. The function $\phi(\cdot)$ is assumed to be smooth and convex with a unique minimum at zero, i.e.,
\[
\phi(0)=0, \quad \phi'(0)=0, \quad \phi''(0)>0
\]
The existence of such constructions follows from results in constructive approximation and smooth function design with prescribed extrema \cite{devore2012} while the link between minima of $h(p)$ and modes of the resulting density is supported by classical critical point theory \cite{morse1934}.

\subsection{Existence of the density and practical implications}
\begin{theorem}[Validity of the Quantile-Density Modulation Construction]
Let $Q_0(p)$ be a baseline quantile function with associated quantile density function
\[
q_0(p) =Q_0'(p)>0,\qquad 0<p<1
\]

Let $h:(0,1)\rightarrow(0,\infty)$ be a continuously differentiable modulation function satisfying assumptions (H1) and (H2). Define
\[
q(p)=q_0(p)h(p)
\]
and
\[
Q(p)=Q(0+)+\int_{0}^{p} q_0(u)h(u),du.
\]

Then $Q(p)$ is a valid quantile function corresponding to a unique absolutely continuous probability distribution.
\end{theorem}

\begin{proof}
Differentiating $Q(p)$ with respect to $p$ gives
\[
Q'(p)=q_0(p)h(p)
\]

Since $q_0(p)>0$ and $h(p)>0$ for all $p\in(0,1)$, it follows that
\[
Q'(p)>0,\qquad 0<p<1
\]

Hence $Q(p)$ is strictly increasing on $(0,1)$ and therefore invertible. Consequently, there exists a unique distribution function
\[
F(x)=Q^{-1}(x).
\]

Moreover, since $Q$ is continuously differentiable, the induced distribution is absolutely continuous. By the standard quantile--density relationship,
\[
f(x)=\frac{1}{Q'(F(x))}
=\frac{1}{q(F(x))}
\]

Therefore, the quantile-density modulation construction generates a valid absolutely continuous probability distribution.
\end{proof}

In general, the probability density function
\[
f(x)=\frac{1}{q(F(x))}
\]
does not admit a closed-form expression.  
Nevertheless, the existence and uniqueness of the density are guaranteed by the absolute continuity of $Q(p)$ \cite{parzen1979}.  All distributional properties including quantiles, moments, simulation and modality are obtained directly from the quantile representation and the results are consistent with modern quantile-based modeling approaches \cite{gilchrist2000, prendergast2019}.
\begin{theorem}[Mode Correspondence Principle]
Let $Q(p)$ be a quantile function with associated quantile density function $q(p)>0$, and let

\[
f(x)=\frac{1}{q(F(x))}
\]

denote the corresponding probability density function.

Then every strict local minimum of $q(p)$ corresponds to a strict local maximum of $f(x)$. Consequently, if $q(p)$ possesses $k$ strict local minima in $(0,1)$, then $f(x)$ possesses $k$ local modes.
\end{theorem}

\begin{proof}
Since (x=Q(p)), we may write

\[
f(x)=\frac{1}{q(p)}
\]

Using the chain rule,

\[
\frac{df(x)}{dx}
= \frac{df/dp}{dx/dp}
= -
 \frac{q'(p)/q^2(p)}{q(p)} 
 =-\frac{q'(p)}{[q(p)]^3}
\]

Hence the critical points of (f(x)) occur precisely when

\[
q'(p)=0
\]

Differentiating again with respect to (x),

\[
\frac{d^2f(x)}{dx^2}= -\frac{q''(p)}{[q(p)]^4}
+
\frac{3[q'(p)]^2}{[q(p)]^5}
\]

At a critical point, where (q'(p)=0), this simplifies to

\[
\frac{d^2f(x)}{dx^2}=
-\frac{q''(p)}{[q(p)]^4}
\]

Therefore, if

\[
q''(p)>0,
\]

then

\[
\frac{d^2f(x)}{dx^2}<0,
\]

which implies that (f(x)) attains a strict local maximum.

which implies that $f(x)$ has a strict local maximum. Thus every strict local minimum of $q(p)$ generates a corresponding local mode of the density function.
\end{proof}

\begin{theorem}[Existence and Control of Bimodality]
Let $Q(p)$ be a quantile function defined on $p \in (0,1)$ of the form
\[
Q(p) = Q(0^+) + \int_0^p q_0(u)\,h(u)\,du
\]
where $q_0(p)$ is a baseline quantile density and $h(p)$ is a strictly positive modulation function. Suppose that $h(p)$ satisfies assumptions \emph{(H1)}–\emph{(H3)} and possesses exactly two local minima at $\xi_1,\xi_2 \in (0,1)$. Then the following hold:
\begin{enumerate}
\item \textbf{Existence.} The probability density function $f(x)$ induced by $Q(p)$ admits at least two local maxima (modes).
\item \textbf{Location (uniform baseline).} If $q_0(p)$ is constant (corresponding to a uniform baseline), the modes of $f(x)$ occur exactly at $x_j = Q(\xi_j)$ for $j=1,2$.
\end{enumerate}
\end{theorem}

\begin{proof}
Let $q(p)=q_0(p)h(p)$ denote the quantile density associated with $Q(p)$.  
Since $Q$ is strictly increasing then the corresponding probability density satisfies
\[
f(x)=\frac{1}{q(p)}, \qquad x=Q(p).
\]
Hence, local maxima of $f(x)$ correspond to local minima of $q(p)$.

The derivative of the quantile density is
\[
q'(p)=q_0'(p)h(p)+q_0(p)h'(p).
\]

\paragraph{Case 1: Uniform baseline.}
If the baseline distribution is uniform, then $q_0(p)=1$ and $q_0'(p)=0$, so that
\[
q'(p)=h'(p).
\]
By assumption \emph{(3)}, $h(p)$ has local minima at $\xi_1$ and $\xi_2$, implying
\[
h'(\xi_j)=0 \quad \text{and} \quad h''(\xi_j)>0, \qquad j=1,2.
\]
Therefore, $q(p)$ attains local minima at $\xi_1$ and $\xi_2$. It follows that
$f(x)$ has local maxima at $x_j=Q(\xi_j)$ for $j=1,2$ establishing both existence
and exact modal location in this case.

\paragraph{Case 2: General baselines.}
For a non-constant baseline quantile density $q_0(p)$, the critical points of $q(p)$
satisfy
\[
h'(p) = -\frac{q_0'(p)}{q_0(p)}\,h(p).
\]
This equation characterizes the locations of the modes in the presence of baseline skewness.For a non-constant baseline quantile density $q_0(p)$, the critical
points of the modified quantile density satisfy

\[
q'(p)=0
\quad \Longleftrightarrow \quad
\frac{h'(p)}{h(p)}
=
-\frac{q_0'(p)}{q_0(p)}.
\]

Thus, the modal structure is determined by the interaction between the
logarithmic derivative of the modulation function and the logarithmic
derivative of the baseline quantile density. When

\[
\left|
\frac{q_0'(p)}{q_0(p)}
\right|
\]

is small relative to

\[
\left|
\frac{h'(p)}{h(p)}
\right|
\]

the critical points of $q(p)$ remain close to the local minima
$\xi_1$ and $\xi_2$ of $h(p)$. Consequently, the parameters
$\xi_1$ and $\xi_2$ remain the primary determinants of the modal
structure, although exact modal locations generally depend on both
$q_0$ and $h$.
\end{proof}
\section{Parameter Estimation}

Let $X_1, X_2, \dots, X_n$ be a random sample from a continuous distribution characterized through its quantile function
\begin{equation}
Q(p; \boldsymbol{\theta}) = Q(0^+) + \int_0^p q_0(u; \boldsymbol{\theta})\, h(u; \boldsymbol{\theta})\, du, \quad 0 < p < 1
\end{equation}
where $q_0(p; \boldsymbol{\theta})$ denotes a baseline quantile density and $h(p; \boldsymbol{\theta})$ is a strictly positive modulation function governing deviations from the baseline and controlling bimodality. The unknown parameter vector is $\boldsymbol{\theta} \in \Theta \subset \mathbb{R}^k$.

Differentiating $Q(p; \boldsymbol{\theta})$ yields the quantile density function
\begin{equation}
q(p; \boldsymbol{\theta}) = \frac{d}{dp}Q(p; \boldsymbol{\theta}) = q_0(p; \boldsymbol{\theta})\, h(p; \boldsymbol{\theta}),
\end{equation}
which determines the shape of the induced distribution.

\subsection{Regularity Conditions and Identifiability}

For the validity of the estimation procedures we have imposed the following regularity conditions:

\begin{itemize}
\item[(C1)] The parameter space $\Theta$ is compact and $\boldsymbol{\theta}_0$ lies in its interior.
\item[(C2)] The functions $q_0(p; \boldsymbol{\theta})$ and $h(p; \boldsymbol{\theta})$ are continuous in both $p \in (0,1)$ and $\boldsymbol{\theta} \in \Theta$, and are continuously differentiable with respect to $\boldsymbol{\theta}$.
\item[(C3)] The modulation function satisfies $h(p; \boldsymbol{\theta}) > 0$ for all $p \in (0,1)$.
\item[(C4)] The quantile function $Q(p; \boldsymbol{\theta})$ is strictly increasing in $p$ for all $\boldsymbol{\theta} \in \Theta$.
\item[(C5)] (Identifiability) If $Q(p; \boldsymbol{\theta}_1) = Q(p; \boldsymbol{\theta}_2)$ for all $p \in (0,1)$, then $\boldsymbol{\theta}_1 = \boldsymbol{\theta}_2$.
\end{itemize}

Condition (C5) ensures that the parameter vector is uniquely determined by the quantile function which is essential for consistent estimation.

\subsection{L-Moment Estimation}

L-moments are linear functionals of the quantile function and provide robust measures of distributional characteristics. The $r$-th population L-moment is defined as
\begin{equation}
\lambda_r(\boldsymbol{\theta}) = \int_0^1 Q(p; \boldsymbol{\theta})\, P_{r-1}(p)\, dp, \quad r \ge 1,
\end{equation}
where $P_{r-1}(p)$ are shifted Legendre polynomials.

Substituting the proposed form of $Q(p; \boldsymbol{\theta})$, we obtain
\begin{equation}
\lambda_r(\boldsymbol{\theta}) 
= \int_0^1 \left[ Q(0^+) + \int_0^p q_0(u;\boldsymbol{\theta}) h(u;\boldsymbol{\theta})\, du \right] P_{r-1}(p)\, dp.
\end{equation}

Interchanging the order of integration yields
\begin{equation}
\lambda_r(\boldsymbol{\theta}) 
= Q(0^+) \int_0^1 P_{r-1}(p)\, dp 
+ \int_0^1 q_0(u;\boldsymbol{\theta}) h(u;\boldsymbol{\theta}) 
\left[ \int_u^1 P_{r-1}(p)\, dp \right] du.
\end{equation}

Let $\hat{\lambda}_r$ denote the sample L-moments computed from the ordered observations. The parameter vector $\boldsymbol{\theta}$ is estimated by solving
\begin{equation}
\lambda_r(\boldsymbol{\theta}) = \hat{\lambda}_r, \quad r = 1, \dots, k,
\end{equation}
or equivalently
\begin{equation}
\hat{\boldsymbol{\theta}}_{LM} = \arg\min_{\boldsymbol{\theta} \in \Theta} 
\sum_{r=1}^k \left[\lambda_r(\boldsymbol{\theta}) - \hat{\lambda}_r \right]^2.
\end{equation}

\begin{theorem}
Under conditions (C1)--(C5), the L-moment estimator $\hat{\boldsymbol{\theta}}_{LM}$ is consistent, i.e.,
\[
\hat{\boldsymbol{\theta}}_{LM} \xrightarrow{p} \boldsymbol{\theta}_0 \quad \text{as } n \to \infty.
\]
\end{theorem}

\begin{proof}
The sample L-moments $\hat{\lambda}_r$ are consistent estimators of $\lambda_r(\boldsymbol{\theta}_0)$ under standard conditions. The mapping $\boldsymbol{\theta} \mapsto \lambda_r(\boldsymbol{\theta})$ is continuous by (C2) and identifiability (C5) ensures a unique solution. The result follows from standard M-estimation arguments.
\end{proof}

\subsection{Least Squares Quantile Estimation}

Let $X_{(i)}$ denote the $i$-th order statistic and define the plotting positions
\begin{equation}
p_i = \frac{i}{n+1}, \quad i = 1, 2, \dots, n.
\end{equation}

Under the assumed model,
\begin{equation}
X_{(i)} \approx Q(p_i; \boldsymbol{\theta}) = Q(0^+) + \int_0^{p_i} q_0(u;\boldsymbol{\theta})\, h(u;\boldsymbol{\theta})\, du.
\end{equation}

The least squares estimator is obtained by minimizing
\begin{equation}
\hat{\boldsymbol{\theta}}_{LS} = \arg\min_{\boldsymbol{\theta} \in \Theta} 
\sum_{i=1}^n \left[
X_{(i)} - Q(0^+) - \int_0^{p_i} q_0(u;\boldsymbol{\theta}) h(u;\boldsymbol{\theta})\, du
\right]^2.
\end{equation}

\begin{theorem}
Under conditions (C1)--(C5), the least squares estimator $\hat{\boldsymbol{\theta}}_{LS}$ is consistent.
\end{theorem}

\begin{proof}
The objective function is a sample analogue of a population criterion based on squared deviations between empirical and theoretical quantiles. Uniform convergence of the empirical quantile process and continuity of $Q(p; \boldsymbol{\theta})$ imply that the criterion converges uniformly to its population counterpart. Consistency follows from standard arguments for nonlinear least squares estimators.
\end{proof}
\section{Bimodality in Quantile Families}

In this section we will demonstrate how the proposed quantile-density modulation framework can be applied to the exponential and logistic distribution families. These families are chosen due to their widespread use in reliability analysis, survival modeling and statistical inference as well as the availability of closed-form quantile representations \cite{johnson1994, kotz2000}.

\subsection{Exponential quantile family}
The exponential distribution with rate parameter $\lambda>0$ is a fundamental model in survival and reliability theory \cite{lawless2003}. Its cumulative distribution function is given by
\[
F(x) = 1 - e^{-\lambda x}, \qquad x>0
\]
and the corresponding quantile function is
\[
Q_0(p) = -\frac{1}{\lambda}\log(1-p) \qquad 0<p<1
\]
The associated quantile probability density function (QPDF) is
\[
q_0(p) = Q_0'(p) = \frac{1}{\lambda(1-p)}
\]

The baseline exponential QPDF is strictly increasing on $(0,1)$ and therefore has no interior local minima. This reflects the unimodal nature of the exponential distribution.

\paragraph{Construction steps:}
The construction of a bimodal exponential-based distribution proceeds as 
\begin{enumerate}
\item Select the baseline exponential quantile function $Q_0(p)$ and its QPDF $q_0(p)$.
\item Choose a modulating function $h(p)$ satisfying assumptions (1)-(3) of Section~2 namely positivity, smoothness and the presence of exactly two local minima.
Here we have used
\[
h(p) = 1 + a \left(
\exp\left(-\frac{(p - p_1)^2}{s^2}\right)
+
\exp\left(-\frac{(p - p_2)^2}{s^2}\right)
\right)
\]
the parameters a and s used here control the shape and strength of the bumps 
\item Define the modified QPDF as
\[
q(p) = q_0(p)\,h(p)
\]
\item Construct the new quantile function by integration:
\[
Q(p) = Q(0^+) + \int_0^p q_0(u)h(u)\,du
\]
\item Since the exponential QPDF has no intrinsic interior minima then the two minima of $h(p)$ directly induce two local minima in $q(p)$ which will later correspond to two modes of the resulting density.
\end{enumerate}
Thus bimodality in the exponential family arises entirely through quantile-density modulation.
\begin{figure}[h]
    \centering
    \includegraphics[width=1\linewidth]{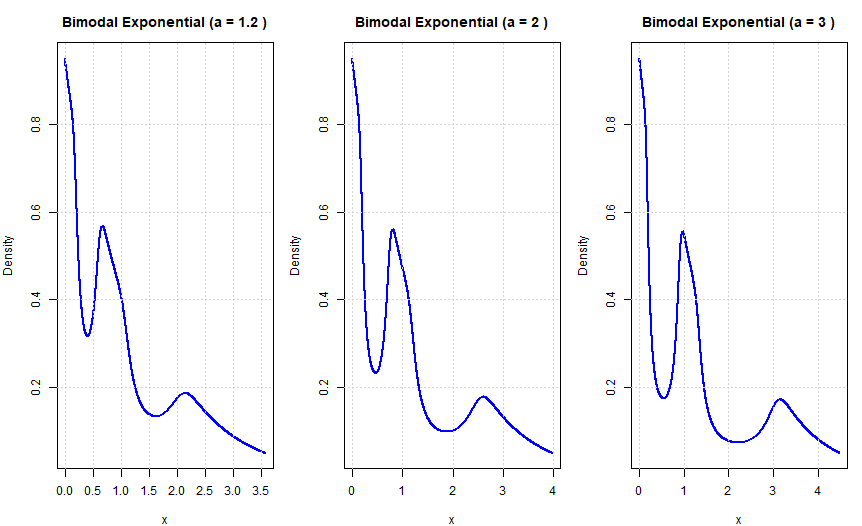}
    \caption{Bimodality in Exponential}
    \label{fig:expo}
\end{figure}
\subsection{Power and Reflected Power Families}

Let $Q_0(p)$ denote the quantile function of the power distribution with parameter $\alpha>0$ and is given by
\[
Q_0(p) = p^{1/\alpha}  \qquad 0<p<1
\]
with corresponding quantile density
\[
q_0(p) = Q_0'(p) = \frac{1}{\alpha}p^{1/\alpha-1}
\]
The power distribution is a special case of the Beta family and is widely used for modeling bounded and skewed data \cite{johnson1995, kotz2000}. Bimodality is introduced by defining the modified quantile density
\[
q(p) = q_0(p)\,h(p) \qquad 0<p<1
\]
\[
h(p) = 1 + a \left(
\exp\left(-\frac{(p - p_1)^2}{s^2}\right)
+
\exp\left(-\frac{(p - p_2)^2}{s^2}\right)
\right)
\]
the parameters a and s used here control the shape and strength of the bumps and $h(p)$ is a strictly positive modulating function possessing exactly two local minima. Under this construction  the resulting probability density will exhibit exactly two modes as local minima of the quantile density correspond to local maxima of the probability density \cite{parzen1979, gilchrist2000}. The support of the resulting distribution remains $(0,1)$.
The reflected power distribution is obtained via the transformation $X^* = 1 - X$ which preserves bounded support while reversing skewness \cite{johnson1995}. The corresponding quantile density is
\[
q^*(p) = q_0(1-p)\,h(1-p) \qquad 0<p<1
\]
which similarly induces bimodality through the two local minima of $h(1-p)$. The reflected power distribution also has support on $(0,1)$.
\begin{figure}[h]
    \centering
    \includegraphics[width=1\linewidth]{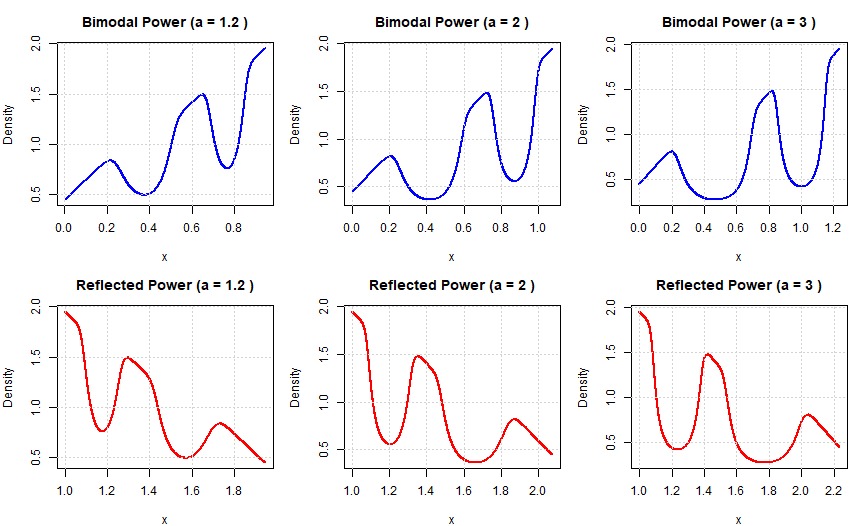}
    \caption{bimodal density for power and reflected power}
    \label{fig:power}
\end{figure}

\subsection{Logistic quantile family}
The logistic distribution is widely used in regression modeling and social sciences due to its symmetry and tractable quantile form \cite{hosmer2013, johnson1995}. The logistic quantile function is given by
\[
Q_0(p) = \mu + \sigma \log\!\left(\frac{p}{1-p}\right) \qquad 0<p<1
\]
where $\mu \in \mathbb{R}$ and $\sigma>0$ are location and scale parameters respectively. The corresponding QPDF is
\[
q_0(p) = Q_0'(p) = \frac{\sigma}{p(1-p)}
\]
Unlike the exponential case the logistic QPDF possesses a unique interior local minimum at $p=0.5$ and the corresponding to the unimodal and symmetric nature of the logistic density.
\paragraph{Construction steps:}
To construct a bimodal distribution based on the logistic family an additional care is required
\begin{enumerate}
\item Select the baseline logistic quantile function $Q_0(p)$ and QPDF $q_0(p)$.
\item Recognize that $q_0(p)$ already contains a local minimum at $p=0.5$.
\item Choose a modulating function $h(p)$ that satisfies assumptions (1)-(3) but is further designed to counteract the intrinsic minimum of $q_0(p)$ at $p=0.5$ while introducing two distinct local minima elsewhere in $(0,1)$.
\item Define the modified QPDF as
\[
q(p) = q_0(p)\,h(p)
\]
\[
h(p) = \exp\left(
\eta
+ \beta \exp\left(-\frac{(p - p_0)^2}{s_0^2}\right)
+ \alpha \left[
\exp\left(-\frac{(p - p_1)^2}{s^2}\right)
+
\exp\left(-\frac{(p - p_2)^2}{s^2}\right)
\right]
\right)
\]
the parameters $\eta$,$\beta$,$\alpha$  and s used here control the shape and strength of the bumps.

\item Construct the resulting quantile function as
\[
Q(p) = Q(0^+) + \int_0^p q_0(u)h(u)\,du
\]
\item Ensure that the combined QPDF $q(p)$ possesses exactly two local minima which correspond to two modes of the induced density.
\end{enumerate}
This construction highlights the importance of balancing the baseline QPDF structure with the modulating function when the underlying family has intrinsic modality.
\begin{figure}
    \centering
    \includegraphics[width=1\linewidth]{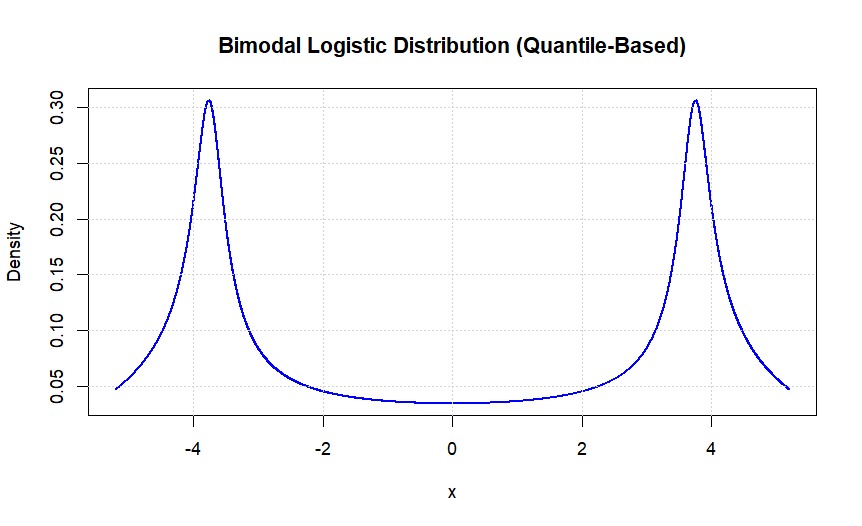}
    \caption{Logistic bimodal density}
    \label{fig:logisticfit}
\end{figure}
\paragraph{Remark}
Baseline distributions whose QPDFs have no interior minima such as the exponential family can admit bimodality solely through quantile-density modulation whereas  distributions with intrinsic QPDF minima such as the logistic family require careful design of the modulating function to ensure that the total number of QPDF minima remains exactly two. In both cases bimodality is entirely determined by the local minima of the modified QPDF and is independent of the existence of a closed-form probability density function.

\subsection*{3.4 GLD Quantile Family}

The Generalized Lambda Distribution (GLD) is a flexible family defined through its quantile representation that allow for a wide range of shapes including skewness and heavy tails\cite{gilchrist2000}
The baseline GLD QPDF is given by
\[
q_0(p) = \frac{1}{\lambda_2} \left( \lambda_3 p^{\lambda_3 - 1} + \lambda_4 (1 - p)^{\lambda_4 - 1} \right)  \quad 0 < p < 1
\]
where \( \lambda_2 > 0 \) is a scale parameter and \( \lambda_3, \lambda_4 \in \mathbb{R} \) control the shape of the distribution. Unlike the logistic family the GLD does not impose a fixed modality structure which makes it particularly suitable for constructing bimodal distributions through external modulation.

\paragraph{Construction steps:}
To construct a bimodal distribution based on the GLD family, the following steps are used:

\begin{enumerate}
\item Select the baseline QPDF \( q_0(p) \).

\item Ensure that \( q_0(p) \) is positive and smooth over \( (0,1) \) that will provide a valid base for transformation.

\item Choose a modulating function \( h(p) \) designed to introduce bimodality while maintaining positivity. In this work, we use
\[
h(p) = \exp\left(
\eta
+ \beta \exp\left(-\frac{(p - p_0)^2}{s_0^2}\right)
+ \alpha \left[
\exp\left(-\frac{(p - p_1)^2}{s^2}\right)
+
\exp\left(-\frac{(p - p_2)^2}{s^2}\right)
\right]
\right)
\]
where \( \eta \in \mathbb{R} \) controls the baseline level \( \beta \in \mathbb{R} \) governs the central peak at \( p_0 \), \( \alpha \in \mathbb{R} \) controls the strength of the side wells (or peaks) and \( s_0, s > 0 \) determine the spread of these features.

\item Define the modified QPDF as
\[
q(p) = q_0(p)\, h(p)
\]

\item The resulting quantile function can be written as-
\[
Q(p) = Q(0^+) + \int_0^p q_0(u)\, h(u)\, du
\]

\item Ensure that the combined QPDF \( q(p) \) will exhibit exactly two local minima which will correspond to only two modes in the induced density function.
\end{enumerate}

This construction highlights the flexibility of the GLD framework achieved by incorporating the modulating function \( h(p) \), using this function one can effectively control the emergence, location and smoothness of bimodality while retaining the desirable distributional properties of the baseline GLD.
\section{Simulation Study for Bimodality}
In this section we have conducted a simulation study to empirically validate the proposed quantile-based bimodality construction with the objective  to demonstrate how bimodality arises through quantile-density modulation and to examine the effect of the modulation parameters on the resulting distributional shape. Random samples were generated using the quantile-based model
\[
Q(p) = \int_0^p q_0(u)\,h(u)\,du
\]
where $q_0(p)$ denotes the baseline quantile density of the generalized lambda distribution (GLD) and $h(p)$ is a strictly positive modulating function. The use of quantile modeling to control distributional shape which follows from the quantile-function modeling paradigm of \cite{parzen1979} and \cite{gilchrist2000} where properties of the probability density are governed by the behavior of the quantile derivative. To induce bimodality in GLD we have constructed the modulating function in log-space as
\[
h(p) = \exp\!\left\{\alpha\left[
\exp\!\left(-\frac{(p-p_1)^2}{s^2}\right)
+
\exp\!\left(-\frac{(p-p_2)^2}{s^2}\right)
\right]\right\}
\]
with $\alpha<0$ and $0<p_1<p_2<1$. This form of the function will ensure strict positivity of the quantile density while introducing two localized depressions which will translate into two modes of the corresponding probability density. Log-transformations are commonly used to guarantee positivity and numerical stability in density and intensity modeling \cite{silverman1986, rue2009}, the Gaussian kernel structure is adopted due to its smoothness and local support which allow precise control over the location depth and width of the induced modal regions.The kernel-based perturbations are widely used in nonparametric density modeling and shape-constrained smoothing \cite{wand1995}. The parameters $\alpha$ and $s$ regulate the strength and smoothness of the bimodality. In this study for each parameter configuration we have generated a large sample using inverse transform sampling using uniformly distributed probabilities and Kernel density estimates. Then the fitted model-based densities were  compared to assess the presence and stability of bimodality.
\begin{figure}
    \centering
    \includegraphics[width=1\linewidth]{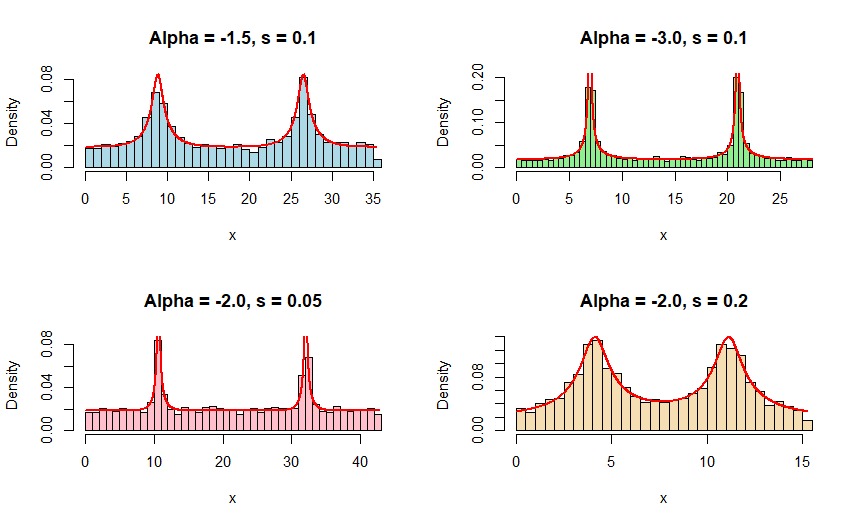}
    \caption{Simulation study of 2-parameter GLD}
    \label{fig:simulation}
\end{figure}
\subsection*{Parameter Estimation}
A simulation study is conducted to evaluate the performance of the proposed quantile least squares estimator for the bimodal GLD model. Data were generated from the model using fixed parameter values $l_1 = 700$, $l_2 = 150$, $l_3 = 0.22$, $l_4 = 0.53$ and $\alpha = -1.13$. The estimation procedure is then applied across a range of sample sizes and the accuracy of the parameter estimates is then assessed using the  mean squared error (MSE). The results are summarized in Table~\ref{tab:simulation}.

\begin{table}[htbp]
\centering
\caption{Estimation Performance of Bimodal GLD Parameters}
\begin{tabular}{c c c c c c}
\toprule
$n$ & Parameter & True Value & Mean Estimate & Bias & MSE \\
\midrule

  & $l1$     & 700.00000 & 700.00000 & 0.00086 & 0.00000 \\
  & $l2$     & 150.00000 & 158.29000 & 8.28790 & 244.96000 \\
200  & $l3$     & 0.22000   & 0.25989   & 0.03989 & 0.00800 \\
  & $l4$     & 0.53000   & 0.51914   & -0.01086 & 0.00608 \\
  & $\alpha$ & -1.13000  & -1.07720  & 0.05285 & 0.06926 \\

\midrule

  & $l1$     & 700.00000 & 700.00000 & 0.00067 & 0.00000 \\
  & $l2$     & 150.00000 & 155.57000 & 5.56790 & 179.69000 \\
500  & $l3$     & 0.22000   & 0.25420   & 0.03420 & 0.00742 \\
  & $l4$     & 0.53000   & 0.52856   & -0.00144 & 0.00447 \\
 & $\alpha$ & -1.13000  & -1.10520  & 0.02479 & 0.02932 \\

\midrule

 & $l1$     & 700.00000 & 700.00001 & 0.00075 & 0.00000 \\
 & $l2$     & 150.00000 & 155.59000 & 5.59010 & 140.51000 \\
1000 & $l3$     & 0.22000   & 0.25568   & 0.03568 & 0.00522 \\
 & $l4$     & 0.53000   & 0.52664   & -0.00336 & 0.00381 \\
 & $\alpha$ & -1.13000  & -1.11370  & 0.01625 & 0.02732 \\

\midrule

 & $l1$     & 700.00000 & 700.00000 & 0.00063 & 0.00000 \\
 & $l2$     & 150.00000 & 153.88000 & 3.88040 & 111.32000 \\
2000 & $l3$     & 0.22000   & 0.25204   & 0.03204 & 0.00458 \\
 & $l4$     & 0.53000   & 0.52801   & -0.00199 & 0.00369 \\
 & $\alpha$ & -1.13000  & -1.12680  & 0.00316 & 0.01963 \\

\bottomrule
\label{tab:simulation}
\end{tabular}
\end{table}

The results have demonstrated that the proposed estimator can perform effectively across all sample sizes and the parameter estimates also remain stable and close to their true values which indicates strong reliability of the estimation procedure.  The shape parameters $l_3$ and $l_4$ can be estimated with small RMSE values which confirms that the model accurately captures the underlying distributional form. In addition, we have seen that the bimodality parameter $\alpha$ has been successfully recovered which have clearly demonstrate the ability of the proposed approach to detect and model complex structural features. Overall, the simulation study have provided a strong evidence that the quantile least squares method offers a robust and effective framework for estimating the parameters of the bimodal GLD model.
\begin{figure}
    \centering
    \includegraphics[width=1\linewidth]{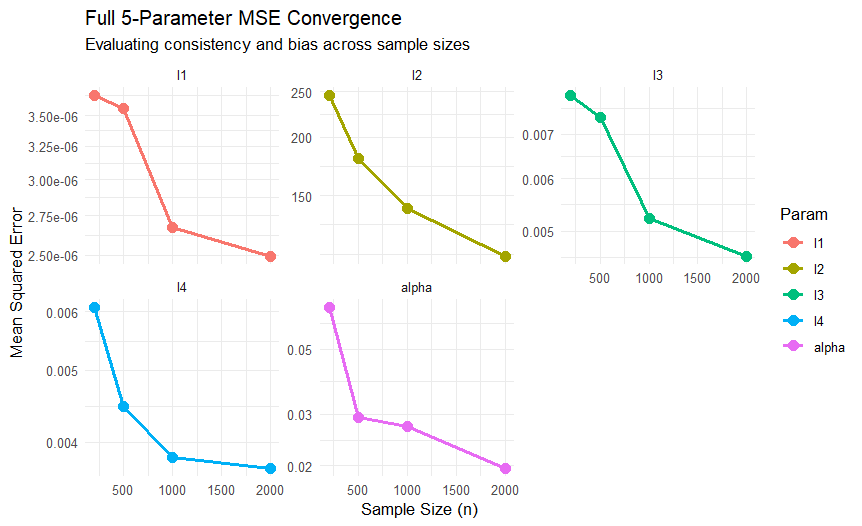}
    \caption{Parameter Estimation using Least Square }
    \label{fig:sim}
\end{figure}
\section{Real-World Application}
To illustrate the practical relevance of the proposed bimodal quantile-based GLD, here we have analyzed a real-world dataset from professional baseball analytics. We have considered the variable \texttt{TEAM\_BATTING\_SO} from the \emph{Moneyball} dataset which is publicly available and has been widely used in applied statistical modeling and sports analytics studies \cite{baumer2015}.This variable records the total number of strikeouts accumulated by each Major League Baseball team over a season. The distribution of strikeouts have shown heterogeneity in team strategy, roster composition and era-specific trends  which make the empirical distribution of data  skewed, heavy-tailed and also exhibits overlapping regimes rather than well-separated modes. These characteristics of this dataset make it quiet suitable to evaluate the flexible distributional models.
Let \(X = \{x_1, \dots, x_n\}\)  which denotes the observed strikeout counts, the visual inspection of the histogram suggests the presence of two partially overlapping strikeout regimes corresponding to lower and higher strikeout team profiles. Rather than modeling the density function directly the proposed approach constructs the distribution through its quantile function \(Q(p)\), where \(p \in (0,1)\) 

The derivative of the quantile function is written as
\[
q(p) = \frac{dQ(p)}{dp}
\]
As a baseline, we adopt the generalized lambda distribution quantile derivative
\[
q_0(p; \lambda_2, \lambda_3, \lambda_4)
=
\frac{1}{\lambda_2}
\left[
\lambda_3 p^{\lambda_3 - 1}
+
\lambda_4 (1 - p)^{\lambda_4 - 1}
\right]
\]
where \(\lambda_2 > 0\) controls scale and \(\lambda_3, \lambda_4 > 0\) govern tail behavior. This formulation corresponds to the FKML parameterization of the GLD \cite{freimer1988}. To introduce bimodality while preserving smoothness and interpretability we have modulated the baseline quantile derivative using a multiplicative function \(h(p)\) which yields
\[
q(p) = q_0(p) \, h(p)
\]
The modulation function is defined as
\[
h(p; \alpha, \epsilon_1, \epsilon_2)
=
\exp \left\{
\alpha \left[
\exp\!\left(-\frac{(p - \epsilon_1)^2}{s^2}\right)
+
\exp\!\left(-\frac{(p - \epsilon_2)^2}{s^2}\right)
\right]
\right\}
\]
where \(\epsilon_1, \epsilon_2 \in (0,1)\) represent the locations of two latent regimes in probability space, \(\alpha < 0\) controls the strength of bimodality and \(s > 0\) determines the smoothness of the modulation.This construction is inspired by kernel-based perturbations which are commonly used in shape-constrained and deformation-based distribution modeling \cite{jones1999, hjort2010} but is different  fundamentally from finite mixture models also the model parameters were estimated via quantile matching. Let \(Q(p)\) denote the  quantile function which is obtained by integrating \(q(p)\) and let \(\widehat{Q}(p)\) denote the empirical quantiles of the observed data. Then the parameters are estimated by minimizing
\[
\sum_{p \in \mathcal{P}} \left[ Q(p) - \widehat{Q}(p) \right]^2
\]
over a dense grid \(\mathcal{P} \subset (0,1)\) 
The resulting quantile function is linearly rescaled to match the observed data range. This step help to preserve the relative shape of the distribution and it ensure the  alignment with the empirical support.
\begin{figure}[H]
            \centering
            \includegraphics[width=1\linewidth]{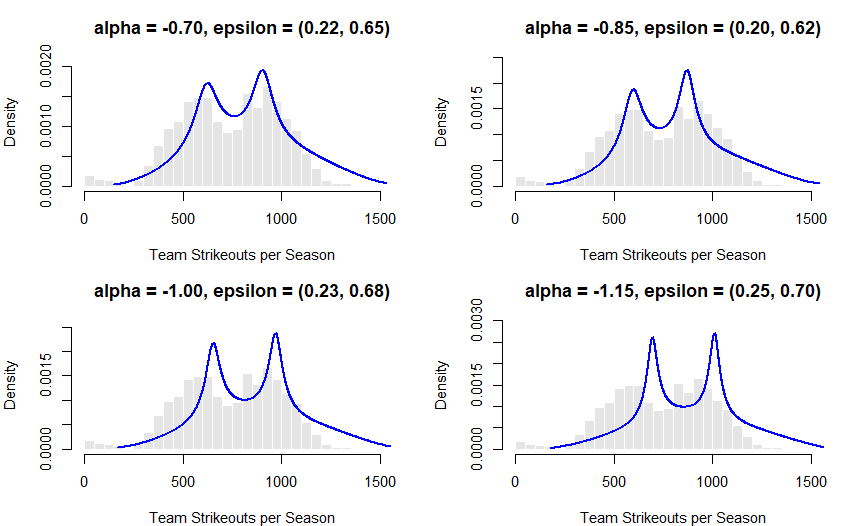}
                \caption{Real world data fitting using bimodal GLD}
    \label{fig:bimodalGLD}
\end{figure}

Figure~\ref{fig:bimodalGLD}  display the fitted densities for four combinations of the bimodality parameters \(\alpha\) and \(\boldsymbol{\epsilon} = (\epsilon_1, \epsilon_2)\). Across all configurations we can see that the fitted distributions can closely track the empirical histogram by remaining smooth and stable in the fitted model. The first mode is controlled by \(\epsilon_1\) that corresponds to lower-strikeout team profiles whereas the second mode is governed by \(\epsilon_2\) which captures the higher-strikeout regimes. When \(\epsilon_2\) is increased it shifts the second regime rightward in a predictable manner while the first regime remains anchored. Also, we can see the two modes overlap substantially which reflects the continuous nature of team performance rather than the sharply separated populations. We have also found that even when the regimes were forced to be close in probability space the fitted densities have exhibit no artificial spikes or oscillations which clearly highlights the robustness of the proposed formulation and make it suitable for real-world datasets where multi-modality is weak or overlapping.

\begin{table}[H]
\centering
\caption{Parameter estimates for the bimodal GLD model using QLS estimators(Moneyball data)}
\label{tab:realdata}
\begin{tabular}{ccccc}
\hline
$l_2$ & $l_3$ & $l_4$ & $\alpha$ & RMSE \\
\hline
151.64 & 0.2223 & 0.5191 & -1.2464 & 31.19 \\
\hline
\end{tabular}
\end{table}
\subsection{Parameter Estimation}
The results presented in Table~\ref{tab:realdata} demonstrate that the proposed bimodal GLD model has provided an effective fit to the Moneyball dataset. The estimated scale parameter $l_2$ have reflected the variability in the data while the shape parameters $l_3$ and $l_4$ have seen to closely capture the underlying distributional form. The bimodality parameter $\alpha$ is negative which indicates the presence of the structural features in the data that deviate from simple uni-modal behavior. This have shown the presence of flexibility in the proposed model in accommodating more complex distributional patterns. Low RMSE values have clearly indicated that the fitted quantile function are well aligned with the empirical quantiles that confirms the accuracy of the estimation procedure. Overall, these results validate the effectiveness of the quantile least squares approach in modeling real-world data using the bimodal GLD framework.

\subsection{Method Validation and Empirical Evaluation}
To effectively validate the proposed framework we have  conducted a series of  experiments that are designed to assess both (i) the reliability of bimodality detection and (ii) the accuracy of parameter recovery under varying distributional regimes. Specifically, we have used three representative cases \textit{deeply separated modes}, \textit{sharp (narrow) modes} and \textit{smooth (overlapping) modes}.

\paragraph{Bimodality Detection}
Bimodality is evaluated using Hartigan’s Dip Test which is a nonparametric statistical test for unimodality which is valid across all experimental conditions as the test produced p-values effectively equal to zero ($p \approx 0$) which leads to  rejection of the null hypothesis of unimodality at any conventional significance level. These result have provided strong evidence that the proposed method  capture bimodal structure irrespective of the degree of separation or overlap between modes.
\begin{table}[h!]
\centering
\caption{Bimodality detection results across regimes.}
\begin{tabular}{lccc}
\hline
\textbf{Regime} & \textbf{Dip Test p-value} & \textbf{Decision} & \textbf{Interpretation} \\
\hline
Deeply Separated Modes & $\approx 0$ & Reject & Strong bimodality \\
Sharp/Narrow Modes & $\approx 0$ & Reject & Distinct but concentrated modes \\
Smooth/Overlapping Modes & $\approx 0$ & Reject & Overlapping yet detectable modes \\
\hline
\end{tabular}
\end{table}

\paragraph{Parameter Recovery Accuracy.}
To quantify estimation performance we have used recovered parameters which are then compared against known ground truth values. Let $\alpha$ denote the primary structural parameter and $s$ be the associated scale parameter whereas the recovery accuracy is measured using relative error in $\alpha$.

\begin{table}[h!]
\centering
\caption{Parameter recovery performance across regimes}
\begin{tabular}{lcccccc}
\hline
\textbf{Regime} & $\alpha_{\text{true}}$ & $\alpha_{\text{est}}$ & \textbf{Error (\%)} & $s_{\text{true}}$ & $s_{\text{est}}$ \\
\hline
Deeply Separated Modes & -3.50 & -3.2390 & 7.46 & 0.08 & 0.0815 \\
Sharp/Narrow Modes & -2.00 & -1.6648 & 16.76 & 0.04 & 0.0419 \\
Smooth/Overlapping Modes & -1.20 & -1.1460 & 4.50 & 0.15 & 0.1552 \\
\hline
\end{tabular}
\end{table}
The results have revealed a clear dependence of estimation accuracy on the geometric characteristics of the underlying distribution. In the deeply separated regime the method has achieved a stable and accurate recovery which clearly reflects the favorable identifiability conditions due to minimal overlap between modes. Conversely, the sharp/narrow regime  contain the most challenging scenario with elevated recovery error. This degradation can be attributed with the increase sensitivity to sampling variation and numerical instability as narrow peaks which induce steep likelihood surfaces and reduced effective sample support. The smooth/overlapping regime have yield the highest estimation accuracy. This behavior suggests that moderate overlap introduces beneficial regularization effects which leads to improved numerical conditioning and more stable optimization trajectories. Across all regimes we can see that the proposed framework have demonstrated consistent and statistically significant bimodality detection which are coupled with robust parameter recovery. The observed error range (4.50\%-16.76\%) is indicating strong overall performance with predictable degradation under extreme sharpness conditions. 
\section{Conclusion}
The Moneyball example have highlighted several advantages of the proposed bimodal quantile-based GLD framework. It not only eliminates the need for explicit mixture weights and latent class assignments but are often easier to interpret and estimate reliability. Here the parameters $\alpha$ and $\boldsymbol{\epsilon}$ provide direct and interpretable control over regime strength and location. Furthermore, the quantile-based formulation naturally accommodate skewness and heavy-tailed behavior which enhance the flexibility of the method in modeling complex data distributions. Overall, the proposed approach offers a coherent and interpretable framework to analyze the real-world data characterized by overlapping regimes where gradual transitions are more realistic than sharply defined subpopulations.
This makes the method particularly suitable for applications in fields such as sports analytics, economics and reliability engineering. Empirical results have shown support for the effectiveness of the approach in both bimodality detection and parameter estimation.
\bibliographystyle{plain}
\bibliography{refrences}

@article{baumer2015,
  author  = {Baumer, Benjamin S. and Kaplan, Daniel T.},
  title   = {OpenIntro statistics: Data sets and stories},
  journal = {Journal of Statistics Education},
  volume  = {23},
  number  = {2},
  year    = {2015}
}

@book{hjort2010,
  editor    = {Hjort, Nils Lid and Holmes, Chris and M{\"u}ller, Peter and Walker, Stephen G.},
  title     = {Bayesian Nonparametrics},
  publisher = {Cambridge University Press},
  year      = {2010}
}

@book{devore2012,
  author    = {DeVore, Ronald A. and Lorentz, George G.},
  title     = {Constructive Approximation},
  publisher = {Springer},
  year      = {2012}
}

@article{freimer1988,
  author  = {Freimer, Marshall and Kollia, Georgia and Mudholkar, Govind S. and Lin, Chun},
  title   = {A study of the generalized Tukey lambda family},
  journal = {Communications in Statistics -- Theory and Methods},
  volume  = {17},
  number  = {10},
  pages   = {3547--3567},
  year    = {1988}
}

@book{fruhwirth2006,
  author    = {Fr{\"u}hwirth-Schnatter, Sylvia},
  title     = {Finite Mixture and Markov Switching Models},
  publisher = {Springer},
  year      = {2006}
}

@book{gilchrist2000,
  author    = {Gilchrist, Warren},
  title     = {Statistical Modelling with Quantile Functions},
  publisher = {Chapman \& Hall/CRC},
  year      = {2000}
}

@article{hartigan1985,
  author  = {Hartigan, John A. and Hartigan, Pamela M.},
  title   = {The dip test of unimodality},
  journal = {Annals of Statistics},
  volume  = {13},
  number  = {1},
  pages   = {70--84},
  year    = {1985}
}

@book{hosking1997,
  author    = {Hosking, Jonathan R. M. and Wallis, James R.},
  title     = {Regional Frequency Analysis: An Approach Based on L-Moments},
  publisher = {Cambridge University Press},
  year      = {1997}
}

@book{hosmer2013,
  author    = {Hosmer, David W. and Lemeshow, Stanley and Sturdivant, Rodney X.},
  title     = {Applied Logistic Regression},
  publisher = {Wiley},
  year      = {2013}
}

@book{johnson1994,
  author    = {Johnson, Norman L. and Kotz, Samuel and Balakrishnan, Narayanaswamy},
  title     = {Continuous Univariate Distributions, Vol. 1},
  publisher = {Wiley},
  year      = {1994}
}

@book{johnson1995,
  author    = {Johnson, Norman L. and Kotz, Samuel and Balakrishnan, Narayanaswamy},
  title     = {Continuous Univariate Distributions, Vol. 2},
  publisher = {Wiley},
  year      = {1995}
}

@book{kotz2000,
  author    = {Kotz, Samuel and Balakrishnan, Narayanaswamy and Johnson, Norman L.},
  title     = {Continuous Multivariate Distributions, Vol. 1},
  publisher = {Wiley},
  year      = {2000}
}

@book{lawless2003,
  author    = {Lawless, Jerald F.},
  title     = {Statistical Models and Methods for Lifetime Data},
  publisher = {Wiley},
  year      = {2003}
}

@article{jones1999,
  author  = {Jones, M. Chris},
  title   = {Simple boundary correction for kernel density estimation},
  journal = {Statistics and Computing},
  volume  = {9},
  pages   = {89--94},
  year    = {1999}
}

@book{mclachlan2000,
  author    = {McLachlan, Geoffrey and Peel, David},
  title     = {Finite Mixture Models},
  publisher = {Wiley},
  year      = {2000}
}

@book{morse1934,
  author    = {Morse, Marston},
  title     = {The Calculus of Variations in the Large},
  publisher = {American Mathematical Society},
  series    = {Colloquium Publications},
  year      = {1934}
}

@article{parzen1979,
  author  = {Parzen, Emanuel},
  title   = {Nonparametric statistical data modeling},
  journal = {Journal of the American Statistical Association},
  volume  = {74},
  number  = {365},
  pages   = {105--121},
  year    = {1979}
}

@article{prendergast2019,
  author  = {Prendergast, Luke A. and Staudte, Robert G.},
  title   = {Quantile-based distributions},
  journal = {Statistical Papers},
  volume  = {60},
  pages   = {327--346},
  year    = {2019}
}

@book{robertson1988,
  author    = {Robertson, Tim and Wright, F. T. and Dykstra, Richard L.},
  title     = {Order Restricted Statistical Inference},
  publisher = {Wiley},
  year      = {1988}
}

@article{rue2009,
  author  = {Rue, H{\aa}vard and Martino, Sara and Chopin, Nicolas},
  title   = {Approximate Bayesian inference for latent Gaussian models by using integrated nested Laplace approximations},
  journal = {Journal of the Royal Statistical Society: Series B},
  volume  = {71},
  number  = {2},
  pages   = {319--392},
  year    = {2009}
}

@article{silverman1981,
  author  = {Silverman, Bernard W.},
  title   = {Using kernel density estimates to investigate multimodality},
  journal = {Journal of the Royal Statistical Society: Series B},
  volume  = {43},
  number  = {1},
  pages   = {97--99},
  year    = {1981}
}

@book{silverman1986,
  author    = {Silverman, Bernard W.},
  title     = {Density Estimation for Statistics and Data Analysis},
  publisher = {Chapman \& Hall},
  year      = {1986}
}

@book{titterington1985,
  author    = {Titterington, D. M. and Smith, Adrian F. M. and Makov, U. E.},
  title     = {Statistical Analysis of Finite Mixture Distributions},
  publisher = {Wiley},
  year      = {1985}
}

@book{wand1995,
  author    = {Wand, Matthew P. and Jones, M. Chris},
  title     = {Kernel Smoothing},
  publisher = {Chapman \& Hall},
  year      = {1995}
}

@article{elal2018,
  author  = {Elal-Olivero, David and G{\'o}mez-D{\'e}niz, Emilio and Conde-S{\'a}nchez, Antonio},
  title   = {A family of bimodal distributions generated by distributions with positive support},
  journal = {Communications in Statistics -- Theory and Methods},
  volume  = {47},
  number  = {23},
  pages   = {5768--5783},
  year    = {2018}
}

@article{gomezdeniz2019,
  author  = {G{\'o}mez-D{\'e}niz, Emilio and Calder{\'i}n-Ojeda, Enrique},
  title   = {Flexible bimodal distributions based on transformations of symmetric distributions},
  journal = {Symmetry},
  volume  = {11},
  number  = {8},
  pages   = {1036},
  year    = {2019}
}

\end{document}